\documentstyle[11pt,Cargesepasp,twoside,epsf]{article}
\reversemarginpar

\begin{document}
\title{Triggered Star Formation in the Scorpius-Centaurus OB Association (Sco OB2)}
 \author{Thomas Preibisch}
\affil{Max-Planck-Institut f\"ur Radioastronomie,
 Auf dem H\"ugel 69, D-53121 Bonn, Germany; preib@mpifr-bonn.mpg.de}
\author{Hans Zinnecker}
\affil{Astrophysikalisches Institut Potsdam, An der Sternwarte 16,
D--14482 Potsdam, Germany; hzinnecker@aip.de}

\begin{abstract}

We explore the star formation history of the Upper Scorpius
OB association, the youngest part of Sco OB2.
A wide field (160 square-degree) survey for low-mass 
pre-main sequence (PMS) stars enabled us to increase the number of known 
low-mass members of Upper Scorpius to nearly 100 stars.
In a detailed analysis of the locations of these stars in the 
HR diagram, taking proper account
of the uncertainties and the effects of unresolved binaries,
we find a mean stellar age of about
5 Myr and no evidence for a significant age dispersion among these
stars.
This implies that the star formation history of the Upper Scorpius
association was dominated by a short star-burst, 
which started about 5 Myr ago and ended probably not more than one or two
 Myr later.
Interestingly, the structure and kinematics of the HI shells surrounding
the Sco OB2 association show that the shock wave of a supernova explosion
in the nearby Upper Centaurus-Lupus association, the oldest part of
Sco OB2, crossed Upper Scorpius just about 5 Myr ago.
This strongly suggests that this supernova shock wave triggered the
star-burst in Upper Scorpius.
\end{abstract}
\keywords{OB Associations; Supernovae; Star Formation History;
Triggered Star Formation; Massive Stars}

\section{Introduction}

OB associations are thought to be the dominant birthplaces for the 
low-mass field star population (Miller \& Scalo 1978; Zinnecker et al.~1993;
Brown et al.~1999). A good knowledge of their
formation history, their stellar content, and their evolution 
is therefore crucial 
for our understanding of the galactic evolution.
Also, OB associations offer a very good opportunity to study the 
origin of the field star initial mass function: they contain the full
range of stellar masses, and since they are very young, their mass function 
can be inferred with minimal corrections for stellar evolution.
Furthermore, the presence of numerous massive stars
creates physical conditions that can be very different from those in 
regions where only low-mass stars form.
The massive stars affect their environment mainly by their ionizing radiation,
their stellar winds, and finally by supernova explosions.
In the immediate neighborhood of the massive stars, these effects are
mostly destructive, since they tend to disrupt the parental molecular cloud.
At somewhat larger distances, however, the shock waves caused by stellar 
winds and supernova explosions can induce collapse of molecular cloud cores 
and thus start star formation, if the right conditions are met
(see Section 5).
Thus the comparison of the stellar content and the star-formation history
of OB associations to those of low-mass star forming regions lacking 
massive stars,  like 
the Taurus-Auriga T association,
can yield important information about the influence of massive stars
on the star formation process: it allows a comparison of
isolated star formation under rather quiescent conditions in T associations
versus clustered star formation in the violent environment of OB associations.

\section{The Sco OB2 association}

\begin{figure}
\plotone{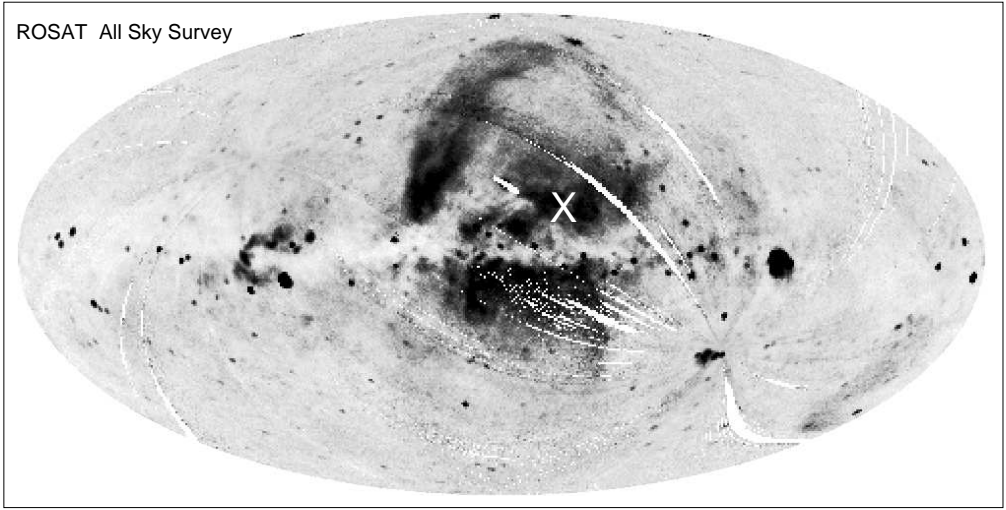}
\caption{All sky map of the diffuse soft (0.75 keV) X-ray emission,
created from the data of the ROSAT All-Sky Survey (Snowden et al.~1995).
The large white {\sf X} symbol slightly above and to the right of the middle
marks the center of the Scorpius-Centaurus association.
The huge (diameter $\sim 120\deg$) roughly spherical feature of enhanced
X-ray emission around Scorpius-Centaurus is caused by the
hot gas in the supernova- and wind-blown bubble
that was created by the massive stars in Sco OB2.
}
\end{figure}
At a distance of only $\sim 140$ pc, the Scorpius-Centaurus association 
is the  OB association
nearest to the Sun.  It contains several hundred B stars
which concentrate in the three subgroups Upper Scorpius (the
youngest subgroup),
Upper Centaurus Lupus (the oldest subgroup), and Lower Centaurus Crux
(cf.~de Zeeuw et al.~1999).
The area is essentially free of dense gas and dust clouds,
probably the consequence of the massive stellar winds
and several supernova explosions, which have cleared the region from
diffuse matter and created a huge system of loop-like H I structures
around the association (cf.~de Geus 1992). The winds and supernova explosions
from the massive stars in Scorpius-Centaurus also created a huge 
bubble of hot gas, that can nicely be seen
in X-ray (see Figure 1) or radio images and constitutes 
the largest structure visible on the sky at these wavelengths.

 The Scorpius-Centaurus association in general, and
Upper Scorpius in particular, was very well investigated by 
the astrometry satellite Hipparcos.
De Zeeuw et al.~(1999) studied the kinematics of the stars in 
Scorpius-Centaurus and could identify 120 members to Upper Scorpius
by their space motions. From this study, 
the population of high-mass ($M_\ast \ga 3\,M_\odot$) members of
Upper Scorpius is completely known. Furthermore, the
ages of these high-mass members are also well known:
De Zeeuw \& Brand (1985) and de Geus et al.~(1989) independently 
derived an age of $5 - 6$ Myr for the B stars in Upper Sco
and found no evidence for a 
significant age spread among these stars.

\section{The low-mass stellar population of Upper Scorpius}

\begin{figure}
\plotone{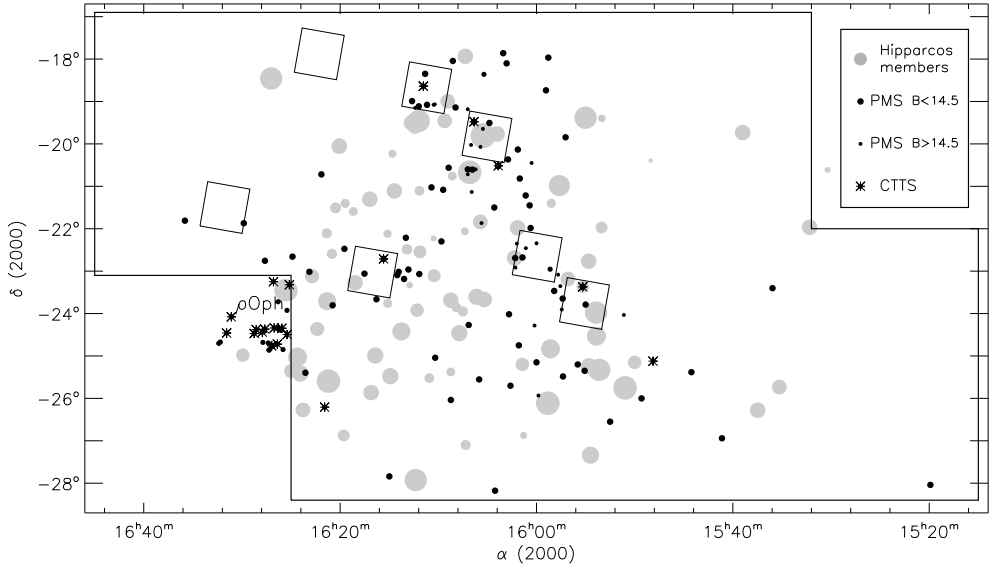}
\caption{Map of our field in Upper Scorpius region, marked by the
thick solid line. The low-mass PMS stars are shown as
solid dots, the Hipparcos high-mass members
as big grey dots with symbol size proportional to the
logarithm of their luminosity. The squares show the fields
investigated by Walter et al.~(1994).}
\end{figure}
While the population of high-mass stars in Upper Scorpius is well known and
investigated, not much was known about the low-mass members until recently.
If the mass function of Upper Scorpius follows the Miller \& Scalo (1979)
field star IMF, one would expect
roughly 1500 low-mass ($M_\star < 2\,M_\odot$) members to be present.
In order to identify a statistically unbiased, magnitude limited sample
of low-mass members,
we have performed a systematic search for PMS stars in Upper Scorpius.
Extending  previous investigations of  small parts of
Upper Scorpius by Walter et al.~(1994) and Kunkel (1999), 
we have surveyed an area of
160 square-degrees, covering the full extent of the association.
By follow-up observations of ROSAT All Sky Survey 
X-ray sources we were able to identify numerous new PMS stars
by their strong lithium absorption lines. This allowed us to increase 
the number of known PMS stars in Upper Scorpius to $\sim 100$ objects
(for all details see Preibisch et al.~1998). 
In Fig.~2 we compare the spatial distribution of the
low-mass members to that of the high-mass members identified by
Hipparcos.
Both populations show a similar,  more or less homogeneous distribution
within a roughly circular area with a radius of some 
$6\deg$ near the center of our field.  

Following the optical characterization of the $\sim 100$ low-mass 
PMS stars, we have placed these stars into the HR diagram (Fig.~3).
The usual way to interpret such a HR diagram is to derive the
 mass and age of each star from its location 
in the diagram by comparison with PMS models. 
However, there are several problems associated with such a procedure
which make the interpretation of the derived ages and masses difficult.
One problem is that any measurement errors, for example uncertainties 
in the photometric data used to determine stellar luminosities, will
cause errors in the derived stellar masses and ages. Also, not all stars
in the association will be at exactly the same distance, adding uncertainty
to the stellar luminosities that have been determined on the adoption
of the mean distance of the association.
Another important problem is that the ages derived from the position
in the HR diagram will 
systematically underestimate the true ages because of the presence
of unresolved binary systems. 
\begin{figure}
\plotone{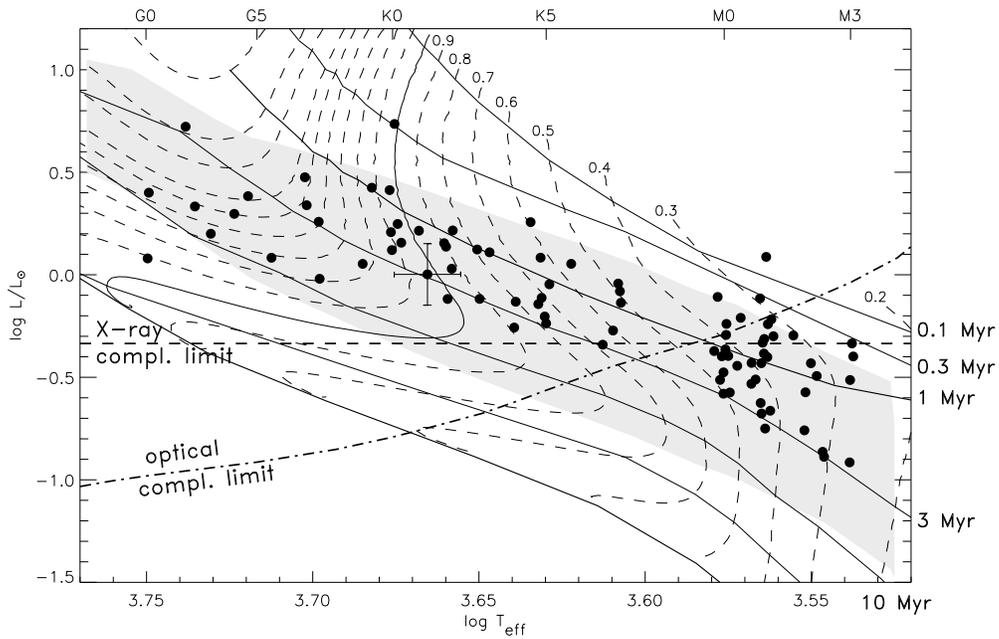}
\caption{HR diagram for the Upper Scorpius PMS stars compared to
the PMS model of D'Antona \& Mazzitelli (1994).
The grey shaded band shows the region in which we expect 90\%
of the PMS stars to lie, based on the assumption of a common age of
5 Myr for all stars and taking proper account of the uncertainties
and the effects of unresolved binaries (for details see Preibisch
\& Zinnecker 1999).}
\end{figure}

We thus decided to use a different approach for our study of the
age structure of the association.
As described in detail in Preibisch \& Zinnecker (1999), we simulated
the location of the members of a model star cluster with a given
age distribution in the HR diagram, taking proper account of the
sources of errors mentioned above. The resulting region in the HR diagram
can then be compared to the observed location of the stars and tell us
whether the underlying assumptions are consistent with the observations
or not.   We find (see Fig.~3) 
that the location of the PMS stars in the HR diagram
is fully consistent with a model assuming that all stars 
have an age of 5 Myr. 
The mean age of the low-mass stars thus 
 agrees very well with the 5 -- 6 Myr previously found for the age of 
the massive stars and shows that low-mass and high-mass stars
are coeval and co-spatial and thus probably have formed together.
The absence of a significant age dispersion (our modeling suggests that
the spread in stellar ages is 
smaller than  about 2 Myr) implies that all stars in the
association have formed more or less simultaneously. This means that the
star-formation process must have started rather suddenly and at the same
time everywhere in the association, and also must have ended rather 
suddenly after at most a few Myr.
The star formation process
in Upper Scorpius thus can be considered as a burst of star formation during
which the stars have been forming much faster than typical for the
quiescent conditions in T associations, where the stars seem to
form more continuously over time.
If all stars in Upper Scorpius have formed within a period of $\la 2 $ Myr,
the star formation rate was
$\psi \ga 1 \times 10^{-3}\,M_\odot/{\rm yr}$.
This is at least one order of magnitude higher than the typical
star formation rate in extended T associations (cf.~Feigelson 1996),
but quite comparable to the star formation rate in the
Orion Nebula cluster (cf.~Hillenbrand \& Hartmann 1998).

To summarize, our results suggest that the star formation history of
Upper Scorpius is characterized by 
\begin{itemize}
\item a nearly simultaneous onset of star formation activity
 in the association about 5 Myr ago, and
\item a rather sudden termination of the star formation activity
 at most a few Myr later.
\end{itemize}
This scenario calls for some
kind of external trigger that initiated the star burst in Upper Scorpius
and some mechanism that terminated the star burst shortly thereafter.
We believe that both effects were caused by
massive stars.

\section{The impact of massive stars on their environment}

Massive stars, upwards of about ten solar masses, profoundly affect 
their environment in several ways (e.g.~Garay \& Lizano 1999).
The most important factors are ionizing radiation, stellar winds,
and supernova explosions.

O-type stars emit intense UV radiation that ionizes and heats 
the surrounding material.
The strong ionizing radiation field next to massive stars can profoundly 
affect nearby cloud cores and the circumstellar material around 
young stellar objects by photoevaporation. This effect
is nicely demonstrated by the proplyds in the Trapezium cluster
(Bally et al.~1998; see also Richling \& Yorke 1998; 2000).
%

Massive stars also have powerful winds that deposit considerable
amounts of momentum and kinetic energy into their surrounding medium.
Typical wind velocities of O-type stars often exceed 1000 km/sec, and
typical values for the mechanical wind power are
$\ga 10^{36}$ erg/sec.
A single O5 star can initiate the disruption of its parental
cloud at a rate $\sim 10^{-2}\,M_\odot/{\rm yr}$ (cf.~Yorke 1986),
i.e.~is perfectly capable of dispersing a $10^4\,M_\odot$ molecular 
cloud completely within only about 1 Myr. 

Finally, after only $\sim 4 - 20$ Myr, the most massive 
($\ga 10\,M_\odot$) stars are expected 
to end their life with  supernova explosions. Each supernova causes a strong 
shock wave that expands with initial velocities of $\ga 10\,000$ km/sec
and transfers typically some $10^{51}$ erg of kinetic energy in the
ambient interstellar medium. 
The shock wave of the supernova will initially expand within the wind-blown
bubble formed by the supernova progenitor. As soon as the supernova blast
wave catches up with the bubble shock front, it will accelerate the expansion
of the bubble (see e.g.~Oey \& Massey 1995 for numerical
evolution models of wind- and supernova-blown bubbles) and
further disrupt the parental molecular cloud (see e.g.~Yorke et al.~1989).

\section{Triggered star formation}

Several recent numerical studies have dealt with the problem of a
shock wave traveling through a molecular cloud and hitting cloud cores
(e.g.~Boss 1995; Foster \& Boss 1996, 1997; Vanhala \& Cameron 1998;
Fukuda \& Hanawa 2000).
These studies found that the outcome of the impact of the shock wave on 
the cloud core mainly depends on the shock velocity:
if the shock wave is faster than about 50 km/sec, the cloud core
is shredded to pieces. If the shock wave is slower than about
10 km/sec, it causes only a slight temporary compression of the core.
Shock waves with velocities in the range of $\sim 10 - 45$  km/sec, 
however, are able to induce collapse of molecular cloud cores.

A potential source of shock waves with velocities in that range
are relatively distant supernova explosions.
The supernova must neither be too
close to the core (because then the shock wave will be too fast and 
destroy the core) nor too far away from the core
(because then the shock wave will be too slow to 
trigger collapse). The appropriate range of velocities that are
suitable to trigger molecular cloud core collapse roughly translates 
into a range of 
distances between 10 pc and 100 pc, the exact values 
obviously depending
strongly on the details of the cloud structure and 
the evolutionary state of the pre-impact core.
Other potential sources of shock waves with velocities in the
desired range include wind-blown bubbles, 
expanding HII regions, and novae.
Another interesting source of shock waves with typical velocities
of a few tens of km/sec are protostellar outflows. Once a burst of
star forming activity is started in a molecular cloud, e.g.~triggered by
a supernova shock wave, many of the forming stars will
produce protostellar outflows and some
of these outflows will hit other cloud cores and can be expected to drive
them into collapse. This process might therefore well be able to
provide a positive feedback to the star formation process and increase
the star formation activity even further.

Several examples for triggered star formation have been discussed in the
literature. For example, it has been suggested that  
star formation in the CMa R1 association (cf.~Herbst \& Assousa 1977)
as well as the Cep OB3 association (cf.~Assousa et al.~1977)
has been triggered by expanding supernova shells.
Oey \& Massey (1995) discuss the dynamics of the superbubble H II region
DEM 152 in the LMC and find evidence for triggered star formation.
In the Cone nebula the
formation of six young stellar objects is thought to be triggered 
by the wind from a B2 star at the center of the system 
(cf.~Thompson et al. 1998).
A good example of the positive feedback on the star forming activity 
provided by protostellar outflows might be the
NGC 1333 star forming region: Knee \& Sandell (2000) investigated the
numerous molecular outflows in NGC 1333 and concluded that apparently
secondary star formation has been and may continue to be triggered
by the outflows of the protostars.

\section{The triggered star burst in Upper Scorpius}

\begin{figure}
\plotone{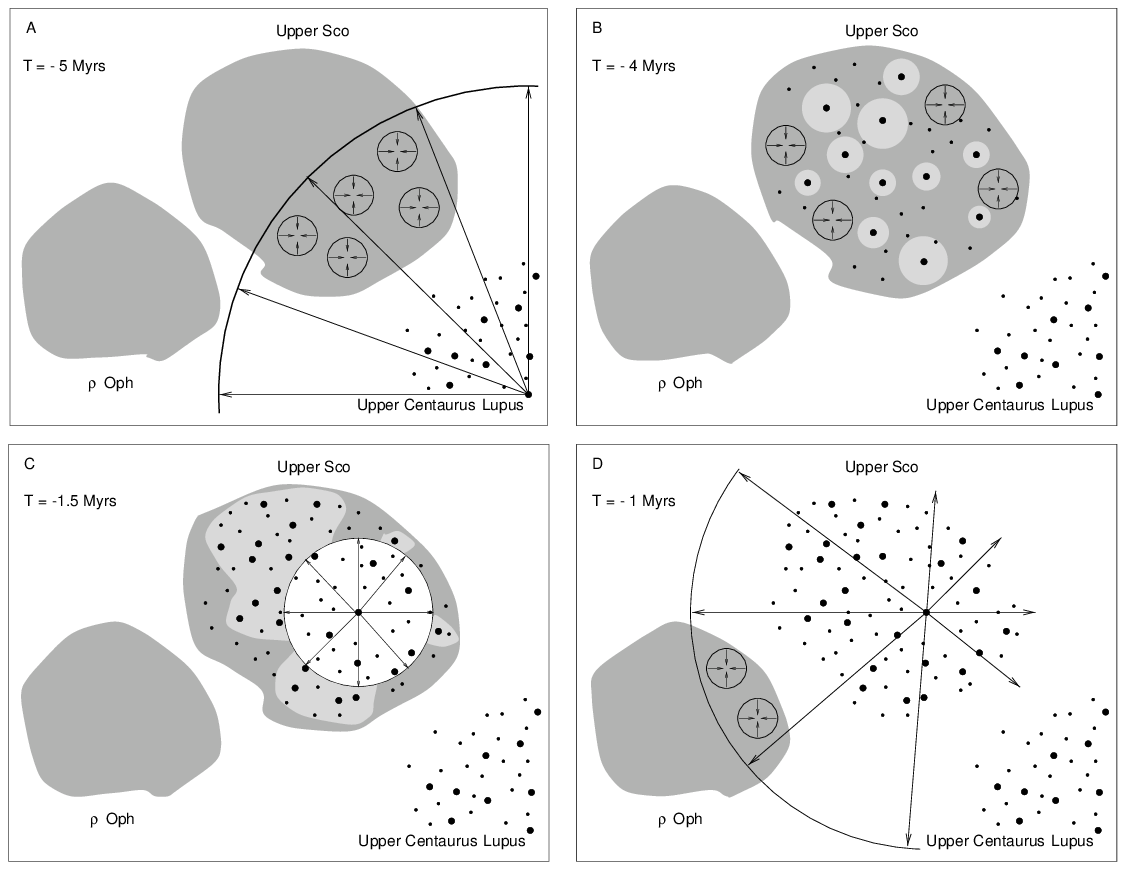}
\caption{Schematic view of the star formation history in the
Scorpius-Centaurus association. Molecular
clouds are shown as dark regions, high-mass and low-mass stars as
large resp.~small dots. For further details see text.}
\end{figure}

In section 3 we concluded that the star burst in Upper Scorpius 
was probably triggered by some external effect.
Interestingly, a suitable trigger is actually available and this allows
us to draw a nice and consistent picture of the star formation history
of Upper Scorpius (see also Fig.~4).
As mentioned above, the Scorpius-Centaurus association
is surrounded by several large H I loops, which were created by
supernova explosions and stellar winds. The kinematic properties
of the largest and thus oldest of these shells 
($v \sim$ 10 km/sec, radius $\sim 110$ pc) 
 suggest that it
was created by a supernova explosion in the Upper Centaurus-Lupus 
association about 12 Myr ago. The geometric and kinematic data suggest
 that this shock wave  passed through the former 
Upper Scorpius molecular cloud just about 5 Myr ago (de Geus 1992). 
This point in time agrees very well with the stellar ages of the
low-mass stars as well as the high-mass stars in Upper Scorpius, that
have been determined in an absolutely independent way.
Furthermore, since the distance from Upper Centaurus-Lupus to Upper Scorpius 
is about 75 pc, this shock wave probably had precisely the properties 
that are required
to induce star formation according to the modeling results mentioned above.
Thus, the assumption that this supernova shock wave triggered
the star formation process in Upper Scorpius (cf.~Fig 4a)
 provides a self-consistent
explanation of all observational data.

During the star burst, probably more than 1000 stars formed within a 
period of at most 1--2 Myr. This suggests that there was a very large
number (several hundreds) of protostellar outflows that might well have
triggered further cloud collapse and accelerated the star formation rate.
However, these outflows presumably also initiated the disruption of 
the molecular cloud. Cloud dispersion must have increased strongly
after a few $10^5$ years, when the new-born massive stars ``turned on''
and started to affect the cloud by their ionizing radiation and their strong 
stellar winds. The combined effect of numerous massive stars probably
affected the cloud so strongly that after a period of $\la 1$ Myr
the star formation process was terminated, simply because all the
remaining cloud material had been disrupted. This explains the narrow
age distribution, i.e.~the lack of stars that formed later than
$\ga 1$ Myr after the onset of the starburst.

Finally, we note that the most massive 
star in Upper Scorpius (presumably a $40\,M_\odot$ star, cf.~de Geus 1992)
exploded as a supernova about 1.5 Myr ago. 
The shock wave  of this explosion fully dispersed 
the remnant cloud material in Upper Scorpius (cf.~Fig.~4c), and must have
reached the $\rho$ Oph cloud within the last 1 Myr (cf.~Fig.~4d).
Interestingly, the $\rho$ Oph cloud actually shows  evidence 
for compression from the rear south-western side and for the presence
of a slow shock (cf.~Motte et al.~1998).
It might well be that this shock wave 
now triggers the current burst of star formation in the $\rho$ Oph 
cloud.

In summary, we conclude that the Scorpius-Centaurus OB association
constitutes one of the best examples of triggered star
formation.

\end{document}